\documentclass[pdftex,12pt,A4paper,superscriptaddress]{revtex4-1}
\usepackage[pdftex]{graphicx}
\usepackage{pifont}
\usepackage{url}
\usepackage{array}
\usepackage{hhline}
\usepackage[normalem]{ulem}
\begin{document}
\newcommand{\circona}{\mbox{\textcircled{\tiny 1}}}
\newcommand{\circonb}{\mbox{\textcircled{\tiny 2}}}
\newcommand{\circonc}{\mbox{\textcircled{\tiny 3}}}
\newcommand{\circond}{\mbox{\textcircled{\tiny 4}}}
\pagenumbering{arabic}\pagestyle{plain}
\title[Double Fano resonances in a composite metamaterial possessing tripod plasmonic resonances]{Double Fano resonances in a composite metamaterial possessing tripod plasmonic resonances}

\author{Y.U. Lee}
\author{E.Y. Choi}
\author{E.S. Kim}
\author{J.H. Woo}
\author{B. Kang}
\author{J. Kim}
\affiliation{Department of Physics and Quantum Metamaterials Research Center, Ewha Womans University, Seoul 120-750, Korea}

\author{Byung Cheol Park}
\author{T.Y. Hong}
\author{Jae Hoon Kim}
\affiliation{Department of Physics, Yonsei University, Seoul 120-749, Korea}

\author{J.W. Wu}
\affiliation{Department of Physics and Quantum Metamaterials Research Center, Ewha Womans University, Seoul 120-750, Korea}


\begin{abstract}
By embedding four-rod resonators inside double-split ring resonators superlattice, a planar composite metamaterial possessing tripod plasmonic resonances is fabricated. Double Fano resonances are observed where a common subradiant driven oscillator is coupled with two superradiant oscillators.
As a classical analogue of four-level tripod atomic system, the extinction spectrum of the composite metamaterial exhibits a coherent effect based on double Fano resonances.
Transfer of the absorbed power between two orthogonal superradiant oscillators is shown to be mediated by the common subradiant oscillator.
\end{abstract}

\maketitle

\section{Introduction}
Control of Fano resonance has been a focus of research in application of artificially structured nano-sized materials
including quantum dots, nano-wires, photonic-crystal slabs, plasmonic nanostructures, as well as metamaterials. \cite{verellen2009fano,8.RevModPhys.82.2257,luk2010fano,rahmani2013fano}
Metamaterials can be designed to possess a broad bright mode and a non-resonant dark mode, which couples each other to exhibit a destructive interference effect such as classical analogue of electromagnetically induced transparency (EIT) phenomenon, corresponding to Fano resonance with the asymmetry parameter equal to zero~\cite{15.PhysRevLett.101.047401,1.nature-material-Stutgart,14.nano-letters-Stutgart1}.

As a further development to engineer metamaterials possessing a higher sensitivity as well as a broad spectral response for sensor applications, multiple Fano resonances are investigated in plasmonic metamaterials.
Multiple Fano resonances can be achieved in two different schemes.
The first scheme is to employ one bright mode and two or more dark modes. Examples are abundant where a bright dipole couples with two dark dipoles or quadrupoles:
one bright dipole resonator and a pair of split ring resonators (SRRs) as the dark elements\cite{liu2012electromagnetically},
double symmetrical U-shaped SRRs with a nanorod between the two SRRs\cite{wang2013double}, plasmonic clusters of heptamer\cite{liu2012multiple} and pentamer\cite{liu2013tuning}, and 3D metamaterials\cite{liu2011three,davis2012analytical,artar2011multispectral}.

The second scheme is to employ one dark mode and two or more bright modes, which already has been utilized in a four-level tripod atomic system to achieve quantum coherent effects such as polarization photon gate\cite{rebic2004polarization}, enhanced cross-phase modulation\cite{li2008enhanced}, and two slow-pulse matching\cite{macrae2008matched}.
%
%
In the four-level tripod atomic system, two different low-lying energy levels $\vert L>$ and $\vert S>$, bright states, are coupled to a common excited state $\vert e>$ by weak probe and trigger lasers,
while the third low-lying energy level $\vert M>$, dark state, is coupled to the excited state $\vert e>$ by a strong pumping laser to provide pathways to both probe and trigger lasers necessary for double EITs.
%
%
While a quantum coherence of $\vert L>$ and $\vert S>$ is maintained, double EIT-based coherent effects are observed.
%
As a classical analogue, a theoretical analysis of Fano correlation effect is reported for a plasmon-exciton-plasmon interaction in an artificial hybrid molecular system, one quantum dot placed in the gap of two identical metal nanoparticles\cite{he2013fano}.
A classical model has already been introduced, where two bright oscillators with different resonance frequencies are coupled to one dark oscillator\cite{bai2012classical}.

\begin{figure}[t]
\begin{center}
   \includegraphics[width=11cm]{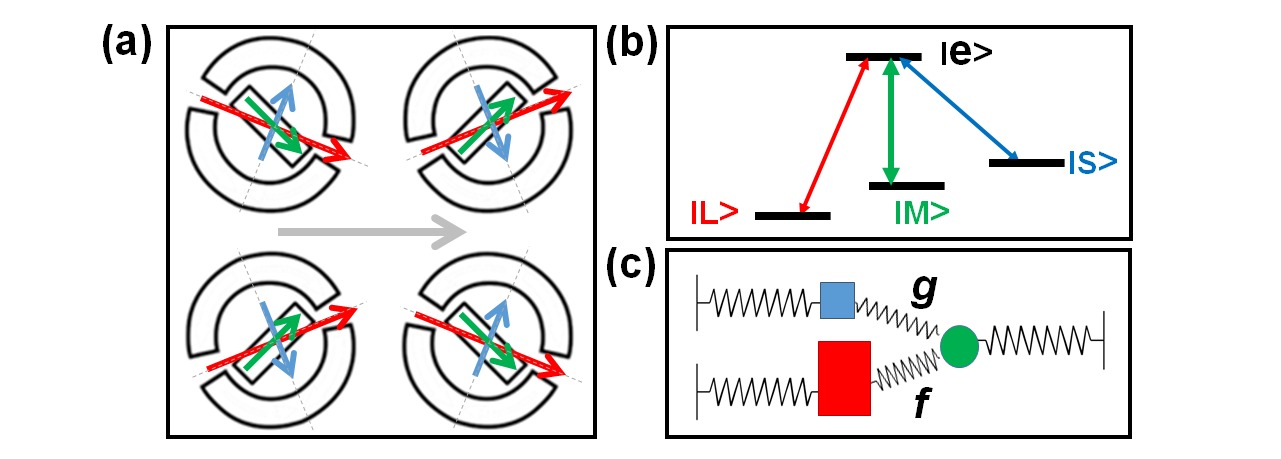}
   \caption
   {\label{fig:coupled-Fano-model}
(a) Schematics of a composite metamaterial is shown with the dipoles in four-rod resonator and double-split ring resonator superlattice for a normally incident wave with horizontal  polarization. Red and blue arrows correspond to two superradiant oscillators, and green arrow correspond to one subradiant oscillator.
(b) Four-level tripod is shown.
$\vert L>$ and $\vert S>$ are two bright states coupled to a common excited state $\vert e>$ by weak probe and trigger lasers,
while the third low-lying energy level $\vert M>$, dark state, is coupled to the common excited state $\vert e>$ by a strong pump laser.
(c) Classical model of one subradiant oscillator, green circle, coupled with two superradiant oscillators, red and blue squares, is shown with coupling strengths $f$ and $g$.
}
   \end{center}
\end{figure}

In this work, we experimentally demonstrate a classical analogue of double Fano resonances in a planar composite metamaterial possessing tripod plasmonic resonances, where a common subradiant dipole corresponding to the dark mode is coupled with two superradient dipoles corresponding to two bright modes.
The composite metamaterial is structured such that four-rod resonators (FRR) are embedded inside double-split ring resonators (DSRR) superlattice.
See Fig.~\ref{fig:coupled-Fano-model}.
Two dipole resonances of DSRR are superradiant modes corresponding to the excitations of $\vert L> \rightarrow \vert e>$ and $\vert S> \rightarrow \vert e>$, and one dipole resonance of FRR is subradiant mode corresponding to the excitation of $\vert M> \rightarrow \vert e>$.
Proximity of the inner diameter of DSRR and the rod-length of FRR permits near-field coherent couplings, and
the composite metamaterial is a tripod metamaterial system possessing two superradiant dipole resonances of DSRR coupled coherently with one common subradiant dipole resonance of FRR in near-field, similar to a tripod atomic system with four atomic levels coupled coherently by coherent optical fields.

Important finding is that double Fano resonances in the composite metamaterial are correlated,
showing up as a transfer of the absorbed power from one superradiant dipole to the other superradient dipole in DSRR.
%

%
In Sec.2, we introduce a planar composite metamaterial possessing tripod plasmonic resonances, and THz extinction spectrum of the fabricated sample is presented along with the finite-difference-time-domain (FDTD) simulated spectrum.
In Sec.3, double Fano resonances are analyzed analytically by a set of coupled equations of motion of two superradiant oscillators and one subradiant oscillators to identify coherent processes of the energy transfer between two superradiant oscillators. In particular, the phase evolutions of superradiant and subradiant oscillators are closely examined.
In Sec.4, the multi-polar nature of near-field coherent interaction is examined by electric field distribution obtained from FDTD theoretical simulation.
Sec.5 summarizes the major findings with conclusion.
%

%
In Appendix the general feature of plasmonic Fano resonance is examined where both superradiant and subradiant oscillators are externally driven.
Analysis based on two coupled oscillators model leads to the fact that the characteristic asymmetric Fano resonance formula of plasmonic structure is kept the same with a modification in the asymmetry parameter $q$.

\section{Sample design, fabrication, and its terahertz spectra}
\subsection{Sample design and fabrication}

\begin{figure}[t]
  \begin{center}
   \includegraphics[width=11cm]{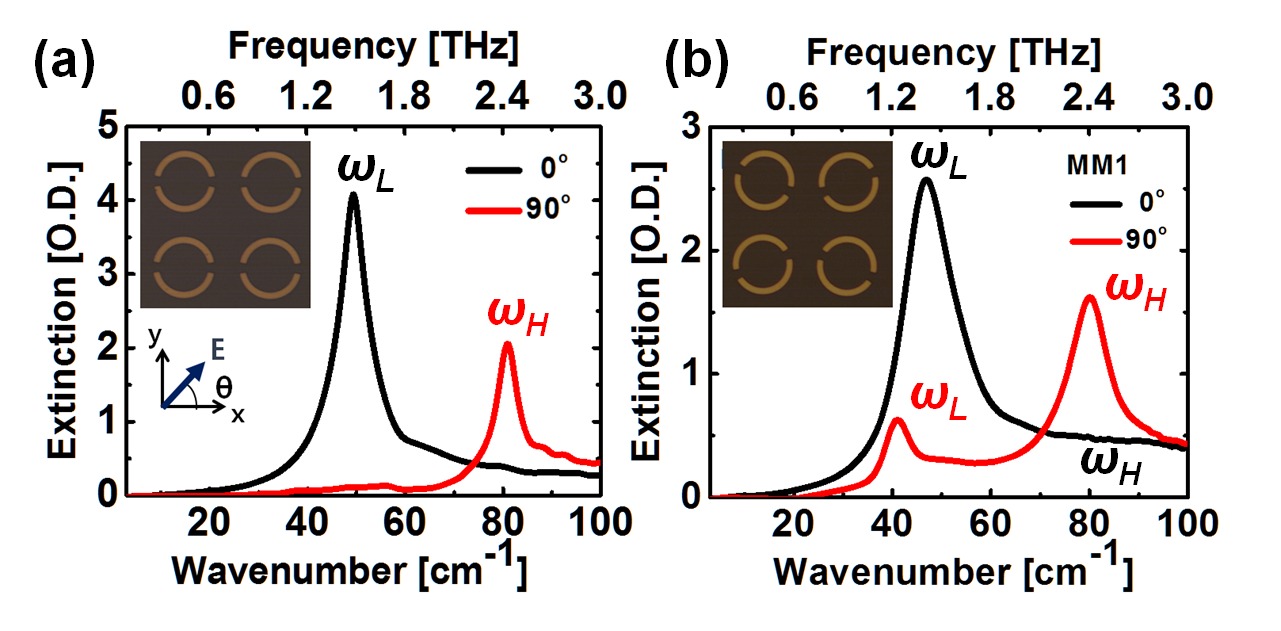}
   \caption
   {\label{fig:S1}
Polarization-dependent optical density plots of measured THz extinction spectra of (a) symmetric superlattice array of double-split ring resonators and (b) asymmetric superlattice array of double-split ring resonators for $x$-(black curve) and $y$-(red curve) polarizations are shown.
}
   \end{center}
\end{figure}

Double split ring resonator (DSRR) is composed of two split rings, and we can construct two different kinds of superlattice arrays of DSRR by arraying them in square lattice either symmetrically or asymmetrically. We compare extinction spectra of symmetric and asymmetric superlattice arrays in Fig.~\ref{fig:S1}, as discussed in \cite{kang2010optical}. When the gap orientation of DSRR is along $x$-axis corresponding to the symmetric superlattice array, two distinct resonances of a low energy $\omega_L$ and a high-energy $\omega_H$ take place in Fig.~\ref{fig:S1}(a). While the high-energy $\omega_H$ resonance is excited for $y$-polarization, the low-energy $\omega_L$ resonance is excited for $x$-polarization. From the extinction spectra of the symmetric superlattice array we find that DSRR exhibits two orthogonal dipolar resonances, a low-energy $\omega_L$ and a high-energy $\omega_H$, which is equivalent to the notion that one DSRR accommodates two independent dipole oscillators with resonances at $\omega_L$ and $\omega_H$\cite{kang2010optical}.

In the asymmetric superlattice array of Fig.~\ref{fig:S1}(b), DSRRs are arrayed in a square lattice such that the gap orientation of DSRR is rotated by angles $\varphi= \pm\frac{\pi}{8}$ alternately with respect to $x$-axis, which allows simultaneous excitations of both low-energy $\omega_L$ and high-energy $\omega_H$ resonances of DSRR upon incidence of a linearly polarized electromagnetic wave. The amount of relative excitations of $\omega_L$ and $\omega_H$ dipole oscillators depends on the angle between the polarization direction and the gap orientation of DSRR\cite{17.JHWoo}.
We pay a particular attention to the extinction spectrum for $x$-polarization, i.e., the black curve in Fig.~\ref{fig:S1}(b). We observe that the high-energy $\omega_H$ is excited as well as the low-energy $\omega_L$ in the asymmetric superlattice array.
Refer to Fig.~\ref{fig:coupled-Fano}(a) for a discussion of the high-energy $\omega_H$ excitation.
In the study of double Fano resonances associated with two bright modes and one dark mode, it is necessary to have simultaneous excitations of two bright modes. In this respect, the asymmetric superlattice array provides an appropriate meta-structure where both high-energy $\omega_H$ and low-energy $\omega_L$ are excited as bright modes for $x$-polarization.
On the other hand, FRR possesses a polarization-independent broad resonance, $\omega_M$, which will be examined more carefully later in connection with double Fano resonances.
\begin{figure}[t]
  \begin{center}
   \includegraphics[width=10cm]{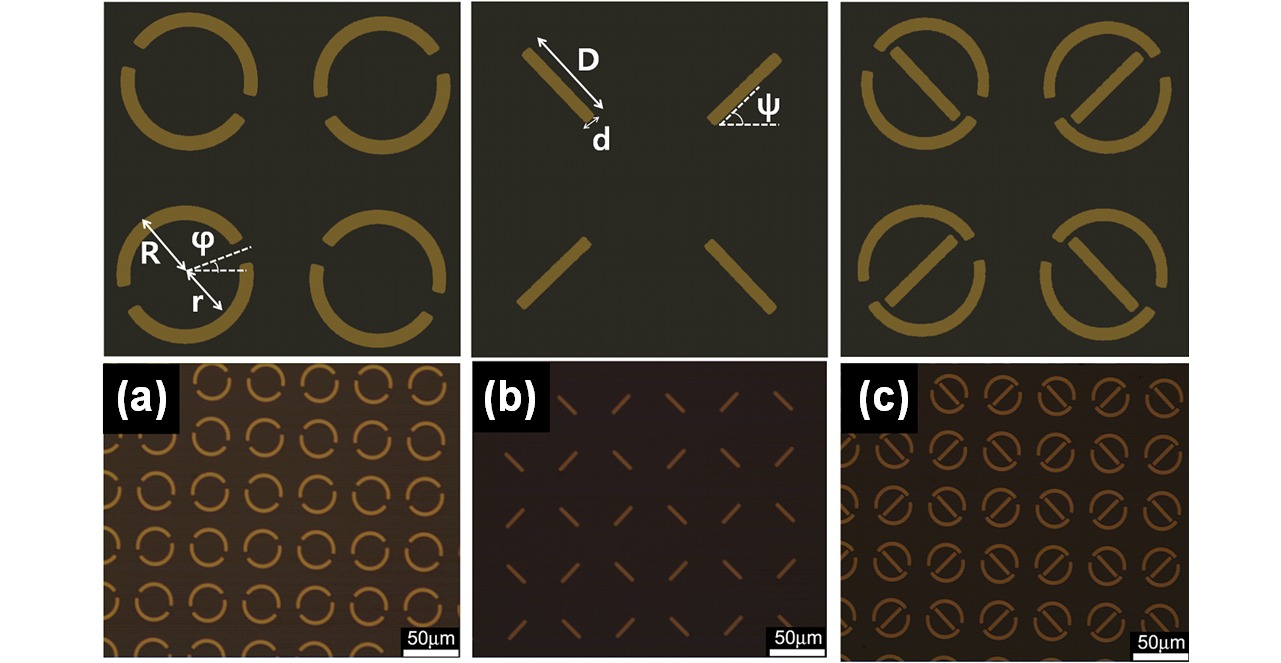}
   \caption
   {\label{fig:sample-microscope}
Optical microscopy images and structural dimensions of (a) double-split ring resonator (DSRR) metamaterial, MM${\bf 1}$,
(b) four-rod resonator (FRR) metamaterial, MM${\bf 2}$, and
(c) composite metamaterial, MM${\bf 3}$ are shown.
$R$=18 $\mu$m, $r$=14 $\mu$m,  $\varphi$= 22.5$^\circ$, and $D$=26 $\mu$m, $d$=4 $\mu$m, $\psi$=45$^\circ$.
}
   \end{center}
\end{figure}

Metamaterials are fabricated by a standard photo-lithography and subsequent lift-off process of Au film with micrometer resolution, which is prepared from e-beam evaporation of 10nm thick titanium adhesion layer and 200nm thick gold layers on 500$\mu$m thick $p$-doped Si wafer.
We call DSRR metamaterial as MM${\bf 1}$, FRR metamaterial as MM${\bf 2}$, and the composite metamaterial as MM${\bf 3}$, respectively, and optical microscopy images are shown in Fig.~\ref{fig:sample-microscope} (a), (b), and (c), with structural dimensions.
MM${\bf 1}$ is an asymmetric superlattice array of DSRRs with the inner radius $r$=14 $\mu$m, the outer radius $R$=18 $\mu$m, the width of circle line
4$\mu$m,  which are arranged with the gap orientation $\varphi= \pm\frac{\pi}{8}$ and the lattice constant 50 $\mu$m.
Previous studies showed that MM${\bf 1}$ exhibits two polarization-dependent dipole resonances with different Q-values\cite{kang2010optical,17.JHWoo,lee2013reflection}.
MM${\bf 2}$ is an array of FRRs with $\psi=\frac{\pi}{4}$, $D$=26 $\mu$m, $d$=4 $\mu$m, and the lattice constant 50 $\mu$m, possessing a 4-fold rotational symmetry. The relative orientation of rods, i.e., $\frac{\pi}{2}$, is set such that the electromagnetic response of MM${\bf 2}$ is the same regardless of the polarization angle of a normally incident electromagnetic wave in an effective medium approximation. That is, MM${\bf 2}$ is isotropic with respect to the polarization direction.
%
%
%
In MM${\bf 3}$, the relative orientation of split-rings in DSRR and rods in FRR results in non-symmetric dipolar interactions between rod of FRR and split-ring of DSRR for the upper and lower arc split-rings.
\subsection{Terahertz spectrum measurement and FDTD simulation spectrum}

It is important to distinguish the electromagnetic characteristics of Au metal in the spectral range of THz and visible regimes. As well-known, in THz frequency regime the real part of dielectric constant of Au has a very large negative value with a small positive imaginary part, which leads that Au behaves as a perfect electric conductor and Au cannot support surface plasmons, electromagnetic surface excitations localized near the surface. However, there exist THz meta-resonances in metamaterials prepared in Au film, originating from an electromagnetic resonance in subwavelength structures, which is called plasmonic resonance\cite{pendry1998low,pendry1999magnetism,pendry2004mimicking,ma2011plasmon,wang2013double, wallauer2013fano}.

Time-domain polarized terahertz transmission measurements are carried out by a Teraview TPS
Spectra 3000 Spectrometer with a resolution of 1.2 cm$^{-1}$ at room temperature in vacuum\cite{teraview}.
Sample size is $10\times10$ $mm^{2}$ and THz beam diameter is about 5$mm$.

FDTD simulations of electromagnetic responses in MM${\bf 1}$, MM${\bf 2}$, and MM${\bf 3}$ are carried out by commercial software FDTD solution Lumerical\cite{lumerical}.
Thickness of Au film in metamaterial samples is 200 $nm$. However, in the FDTD simulation we adopted the thickness of Au film as $t$ = 4 $\mu m$, which satisfies the relation $t \ll  \lambda$, with the setting of an override maximum mesh step size of dz = 1 $\mu m$ and minimum mesh step size 0.00025 $\mu m$. As far as the relation $t \ll  \lambda$ holds, the thickness of Au is not relevant, since Au film itself does not support any resonances and THz resonances originate from the subwavelength structure of metamaterial\cite{lumerical,chen2006active,fdtd}.

A unit cell of metamaterial has been set as the unit cell of FDTD simulation and the periodic boundary conditions are adopted in $x$- and $y$-directions and the perfectly matched layer boundary condition is adopted in $z$-direction.
In order to identify whether the broad high-energy $\omega_H$ resonance of MM${\bf 1}$ stems from a coupling between neighboring DSRRs, we examined FDTD simulated extinction spectrum by varying the horizontal distance between DSRRs. And we found that there always exists the broad high-energy $\omega_H$ resonance with the identical spectral shape except for a decrease in the magnitude of extinction simply owing to a reduced number of meta-particles per unit area.

For the dielectric permittivity of $p$-doped silicon substrate we employed the values obtained from THz spectrum measurement, and the dielectric permittivity of Au is adopted from CRC handbook \cite{weast1988crc}, which gives a typical value of $\epsilon = -4910 + i 4.915\times 10^{-5}$ at 2.0 THz.

\begin{figure}[t]
  \begin{center}
   \includegraphics[width=12cm]{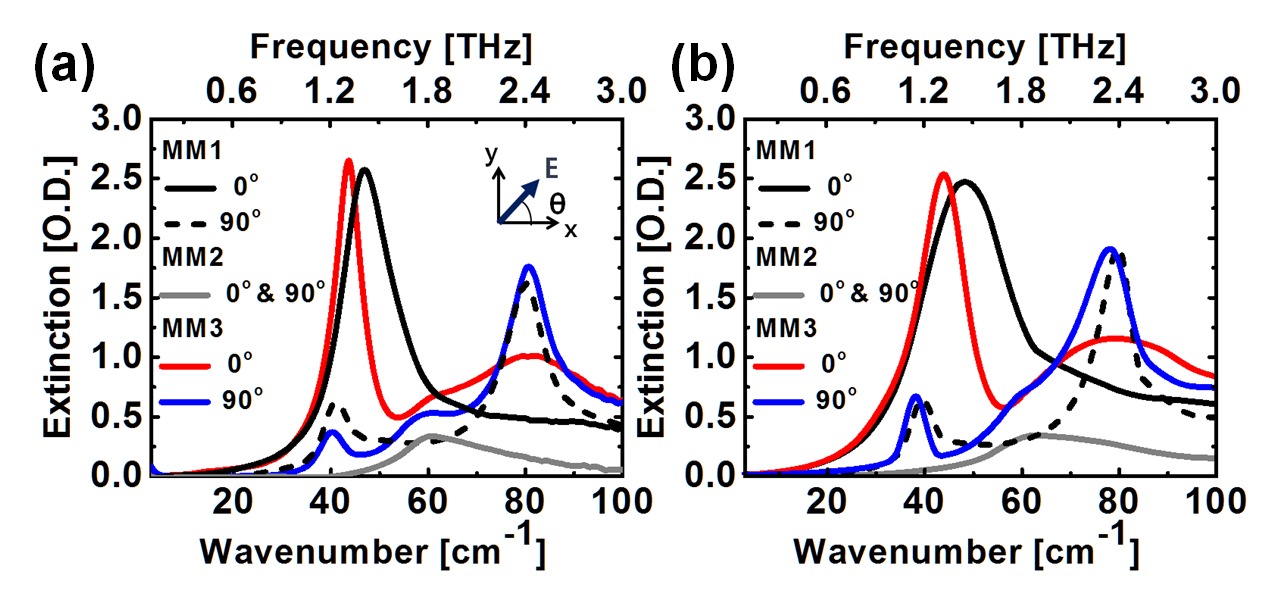}
   \caption
   {\label{fig:S2}
(a) Polarization-dependent optical density plots of measured THz extinction spectra of MM${\bf 1}$, MM${\bf 2}$, and MM${\bf 3}$ are shown at $x$- ($\theta = 0^\circ$) and $y$- ($\theta = 90^\circ$) polarizations. (b) Polarization-dependent optical density plots of FDTD simulated spectra are shown for  MM${\bf 1}$, MM${\bf 2}$, and MM${\bf 3}$.
}
   \end{center}
\end{figure}

Figure~\ref{fig:S2} shows the polarization-dependent optical density plots of (a) measured and (b) simulated THz extinction spectra of MM${\bf 3}$ for a normally incident wave with $x$-polarization, $\theta=0^{\circ}$, (red curve) and $y$-polarization, $\theta=90^{\circ}$, (blue curve). Also shown are the extinction spectra of MM${\bf 1}$ with $x$-polarization, $\theta=0^{\circ}$, (solid black curve) and $y$-polarization, $\theta=90^{\circ}$, (dashed black curve) and the extinction spectrum of MM${\bf 2}$ (gray curve).
MM${\bf 2}$ exhibits polarization-independent broad resonance (gray curve), and
the composite metamaterial MM${\bf 3}$ shows polarization-dependent resonances (red and blue curve), which is associated with polarization-dependent meta-resonances of MM${\bf 1}$.
Comparison of Fig.~\ref{fig:S2}(a) and (b) shows that there is an excellent agreement between measured and simulated extinction spectra.

\begin{figure}[t]
  \begin{center}
   \includegraphics[width=12cm]{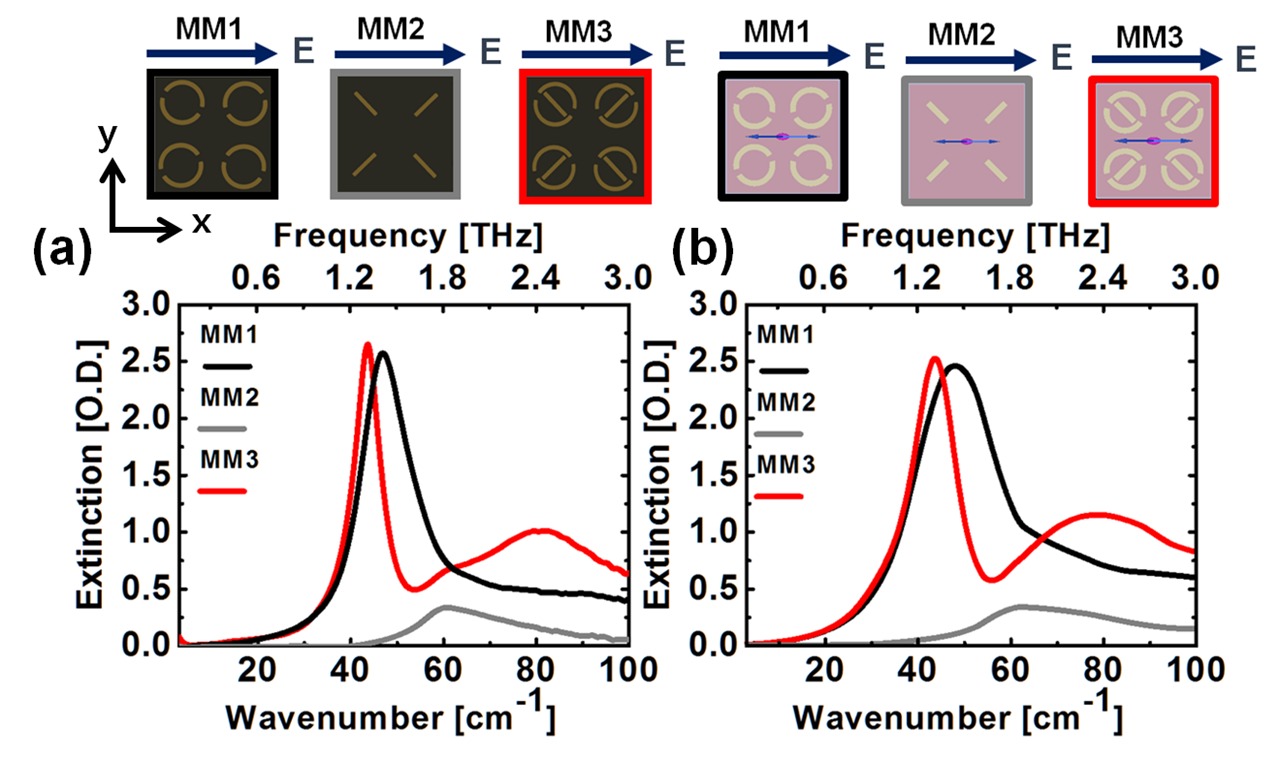}
   \caption
   {\label{fig:measurement Fano}
(a) Optical density plots of measured THz extinction spectra of MM${\bf 1}$ (black), MM${\bf 2}$ (gray), and MM${\bf 3}$ (red) are shown  for $x$-polarization. (b) Optical density plots of FDTD simulated spectra are shown for MM${\bf 1}$ (black), MM${\bf 2}$ (gray), and MM${\bf 3}$ (red).
}
   \end{center}
\end{figure}

\begin{figure}[t]
  \begin{center}
   \includegraphics[width=9cm]{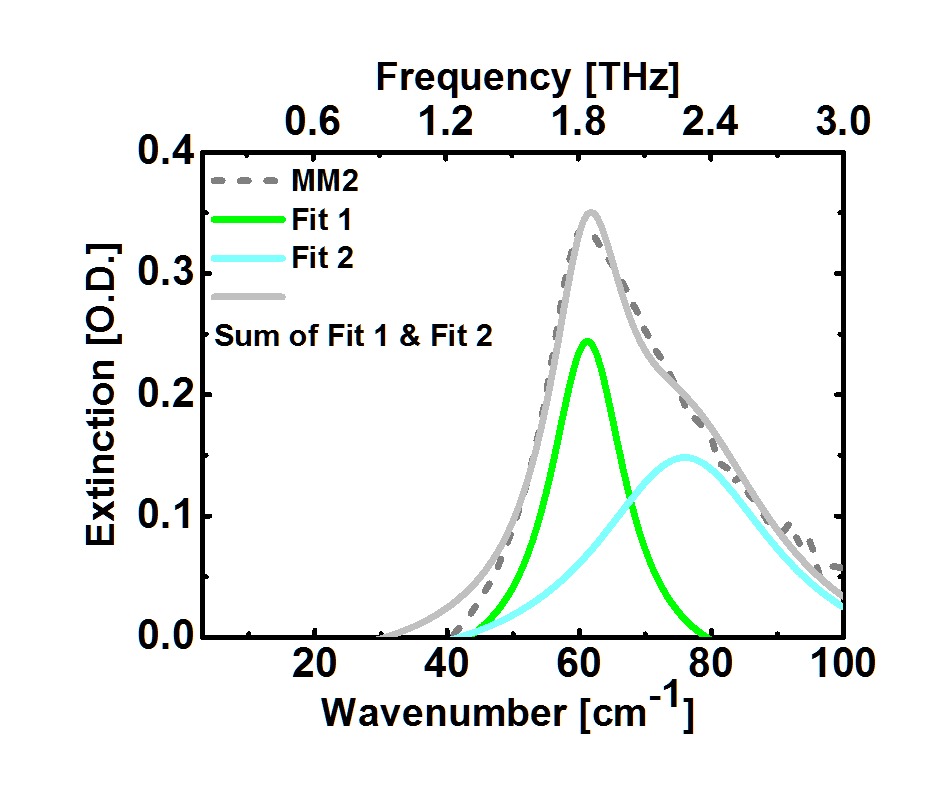}
   \caption
   {\label{fig:S3}
MM${\bf 2}$ extinction spectrum is deconvoluted to 2 Lorentzian resonances. Linear sum of one resonance ${\omega_{M}}^{(1)} = 60 {\rm cm}^{-1}$ and ${\gamma_{M}}^{(1)} = 14 {\rm cm}^{-1}$ with $Q^{(1)} = 4.4 $ (green curve) and  the other resonance ${\omega_{M}}^{(2)} = 76 {\rm cm}^{-1}$ and ${\gamma_{M}}^{(2)} = 34 {\rm cm}^{-1}$ with $Q^{(2)} = 2.2 $ (blue curve) is shown in the solid gray curve along with the measured extinction spectrum of MM${\bf 2}$ (dashed gray curve).
}
   \end{center}
\end{figure}

In the study of double Fano resonances taking place in MM${\bf 3}$, we closely examine the extinction spectra with $x$-polarization, $\theta=0^{\circ}$, which are re-plotted in Fig.~\ref{fig:measurement Fano}.
As a digression we deconvolute the extinction spectra of MM${\bf 2}$ to 2 Lorentzian resonances.
Figure~\ref{fig:S3} shows how the extinction spectrum of MM${\bf 2}$ is composed of Lorentzian resonances.
We find that the broad extinction spectrum of MM${\bf 2}$ is a linear sum of one resonance ${\omega_{M}}^{(1)} = 60 {\rm cm}^{-1}$ with a width ${\gamma_{M}}^{(1)} = 14 {\rm cm}^{-1}$ corresponding to $Q^{(1)} = 4.4 $ (green curve) and the other resonance ${\omega_{M}}^{(2)} = 76 {\rm cm}^{-1}$ with a width ${\gamma_{M}}^{(2)} = 34 {\rm cm}^{-1}$ corresponding to $Q^{(2)} = 2.2 $ (cyan curve). Comparison of the sum of 2 Lorentzian resonances (solid gray curve) and the measured extinction spectrum (dashed gray curve) shows a good agreement. As will be discussed later, the Lorentzian resonance ${\omega_{M}}^{(1)} = 60 {\rm cm}^{-1}$ takes part in giving rise to double Fano resonances, while the Lorentzian resonance ${\omega_{M}}^{(2)} = 76 {\rm cm}^{-1}$ stays as a background absorption since the spectral location of resonance is far away from bright mode resonances of MM${\bf 1}$. Refer to Fig.~\ref{fig:coupled-Fano} and its discussion as well as Tab.~\ref{table-isolate}.

Let's examine the characteristics of MM${\bf 1}$ and MM${\bf 2}$ spectra in Fig.~\ref{fig:measurement Fano}(a).
MM${\bf 1}$ (black) possesses one low-energy resonance with a large oscillator strength at $\omega_L$ = 1.4THz and the other high-energy resonance with a small oscillator strength stretching with $\omega_H$ spanning from 1.8THz to 3.0THz. On the other hand, MM${\bf 2}$ (gray) possesses a broad resonance which is a linear sum of ${\omega_{M}}^{(1)}$ = 1.8THz and ${\omega_{M}}^{(2)}$ = 2.3THz.

Comparison of the linear absorption spectra of MM${\bf 1}$ and MM${\bf 2}$ shows two notable features.
First, the total absorption cross section of MM${\bf 1}$ is much larger than  that of MM${\bf 2}$, indicating that the dipolar interaction with an  incident electric field is much stronger in MM${\bf 1}$ than in MM${\bf 2}$.
Therefore, MM${\bf 3}$ is composed of a strongly driven MM${\bf 1}$ oscillator and a weakly driven MM${\bf 2}$ oscillator with the two interacting by a near-field coupling.
Second, the entire spectrum of MM${\bf 2}$ is contained under the envelope of MM${\bf 1}$ spectrum, with ${\omega_{M}}^{(1)}$ resonance in  MM${\bf 2}$ positioned between $\omega_L$ resonance and $\omega_H$ resonance in MM${\bf 1}$.
Next, we examine the difference in extinction spectrum between MM${\bf 1}$ and MM${\bf 3}$ qualitatively by paying attention to the low-frequency and high-frequency peaks in Fig.~\ref{fig:measurement Fano}(a).
At the low-frequency peak, MM${\bf 3}$ (red) spectrum shows a dip at $\approx$ 1.6THz, the spectral position being on the blue side of MM${\bf 1}$ (black) low-energy resonance $\omega_L$ = 1.4THz.
On the other hand, at the high-frequency peak, the extinction is enhanced almost two-fold at 2.4THz in MM${\bf 3}$ (red) spectrum  when compared with MM${\bf 1}$ (black) spectrum.
These features at low- and high-frequency peaks show that there exists a correlation between changes in the low-frequency and high-frequency peaks. We note that the spectral positions of dip at $\approx$ 1.6THz and enhancement in the extinction at 2.4THz  are located on the red and blue sides of MM${\bf 2}$ (gray) resonance ${\omega_{M}}^{(1)}$=1.8THz, respectively.

\section{Double Fano resonances}
\subsection{Correlation of double Fano resonances}
Now we examine the correlation between changes in the low-frequency and high-frequency peaks in terms of double Fano resonances analytically.
Let's call the superradiant dipole with $\omega_L$ resonance and the superradiant dipole with $\omega_H$ resonance in DSRR as $u(t)$ and $v(t)$, respectively, and the subradiant dipole with ${\omega_{M}}^{(1)}$ resonance in FRR as $\zeta(t)$.
In the composite metamaterial MM${\bf 3}$, DSRRs are arrayed with the orientation angle $\varphi= \pm\frac{\pi}{8}$, and upon normal incidence of $x$-polarization wave, both $u(t)$ and $v(t)$ are excited.
%

Noting that a pair of superradiant dipoles at the upper and lower arcs of DSRR interact with subradiant dipole of FRR symmetrically, dipolar interactions between the superradiant dipoles $u(t)$ \& $v(t)$ and the subradiant dipole $\zeta(t)$ are expressed as %
\begin{equation}
U = - f\Omega_r^2 u(t)\zeta(t) -  g\Omega_r^2 v(t)\zeta(t)
 {\label{eq:dipole}}
\end{equation}
where $f$ and $g$ are the coupling strengths of $u(t)$ and $v(t)$ with $\zeta(t)$, respectively.
$\Omega_r$ is the frequency associated with the characteristic coherent coupling
between superradiant dipoles and subradiant dipole.
%

As shown in Appendix, an asymmetric Fano lineshape takes place even when the subradiant oscillator is externally driven. In the composite metamaterial MM${\bf 3}$, FRR resonator  $\zeta(t)$ corresponds to the externally driven subradiant oscillator.
In double Fano resonances we have two superradiant oscillators, and from Eq.~(\ref{eq:b}) and ~(\ref{eq:d}) in Appendix the coupled equations of motion are straightforward with superradiant $u(t)$ and superradiant $v(t)$ and subradiant $\zeta(t)$ externally driven by a normally incident $x$-polarization wave.

\begin{equation}
\frac{d^2 u(t)}{dt^2} + \gamma_u \frac{du(t)}{dt} + \omega_u^2 u(t) -f\Omega_r^2 \zeta(t)
= \cos(\frac{\pi}{8})e^{-i\omega_s t}
 {\label{eq:x}}
\end{equation}
\begin{equation}
\frac{d^2 v(t)}{dt^2} + \gamma_v \frac{dv(t)}{dt} + \omega_v^2 v(t) -g\Omega_r^2 \zeta(t)
= \sin(\frac{\pi}{8}) e^{-i\omega_s t}
{\label{eq:y}}
\end{equation}
\begin{equation}
\frac{d^2 \zeta(t)}{dt^2} + \gamma_\zeta \frac{d\zeta(t)}{dt} + \omega_\zeta^2 \zeta(t)
-\Omega_r^2 \{u(t)f+v(t)g\}
= \eta e^{-i\omega_s t}
{\label{eq:dark}}
\end{equation}
Note that the subradiant dipole $\zeta(t)$ is externally driven with a driving amplitude of $\eta$.

In the absence of near-field couplings with the subradiant oscillator $\zeta(t)$, i.e., $f=g=0$, two superradiant oscillators $u(t)$ and $v(t)$ are excited independently with the driving amplitudes of $\cos(\frac{\pi}{8})$ and $\sin(\frac{\pi}{8})$, respectively.
When near-field couplings of $u(t)$ and $v(t)$ with $\zeta(t)$ are turned on, i.e., $f\ne 0$ and $g\ne 0$, there occurs an interplay between $u(t)$ and $v(t)$ through coupling $\Omega_r^2$.
That is, $\omega_L$ oscillator $u(t)$ and $\omega_H$ oscillator $v(t)$ reside in one and the same DSRR, and Fano resonances in $u(t)$ and $v(t)$ are correlated. This is similar to the control of EITs in a tripod atomic system where the absorption spectrum of the hyperfine field shows a dependence on the powers of the Zeeman EIT signal\cite{macrae2008matched}. A classical example is to control the absorbed power spectrum of one metal nanoparticle by manipulating the distance between quantum dot and the other metal nanoparticle in a system composed of one quantum dot placed in the gap of two identical metal nanoparticles\cite{he2013fano}.

%
%
Analytical expression of double Fano resonances by the coupled equations of motion Eq.~(\ref{eq:x})-(\ref{eq:dark}) allows to identify how the two independent orthogonal superradiant oscillators interact indirectly via near-field couplings with the common subradiant oscillator.

\begin{figure}[t]
\begin{center}
\includegraphics[width=12cm]{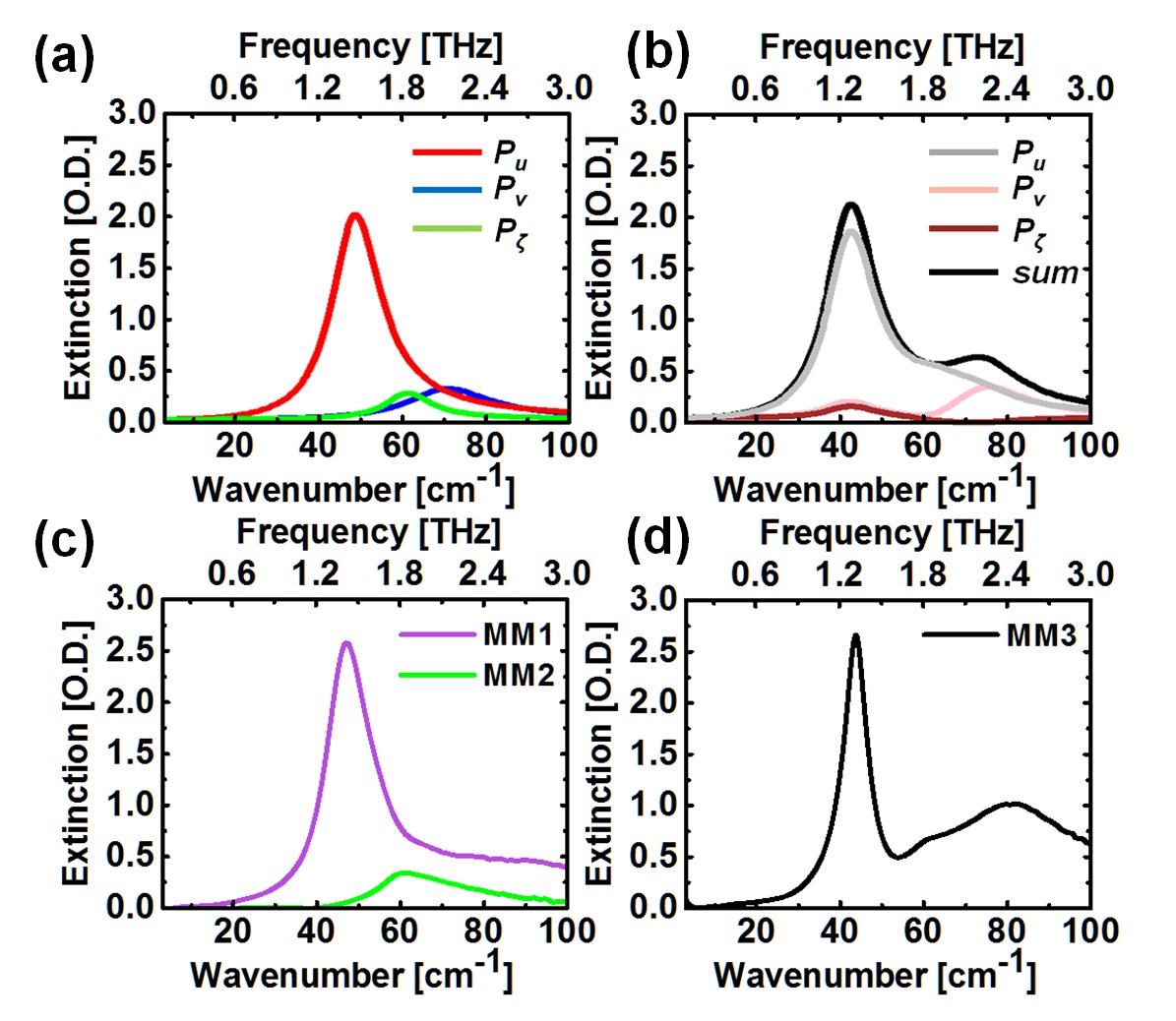}
  \caption
   {\label{fig:coupled-Fano}
The extinction spectra of absorbed powers are plotted.
In (a), red, blue, and green curves  correspond to the absorbed powers $P_x(\omega_s)$ of isolated dipoles $x(t)$, $x=[u,v,\zeta]$ obtained from the coupled Eq.~(\ref{eq:x})-(\ref{eq:dark}) with $f=g=0$.
In (b), gray, pink, and dark brown curves correspond to the absorbed powers $P_x(\omega_s)$ of coupled dipoles $x(t)$ with $f\ne 0$ and $g\ne 0$, respectively.
Black curve in (b) is the sum of gray, pink, and dark brown curves.
In (c) and (d), purple, green, and black curves are the measured extinction spectra of MM${\bf 1}$, MM${\bf 2}$, and MM${\bf 3}$ re-plotted from Fig.~\ref{fig:measurement Fano}(a).}
\end{center}
\end{figure}

In Fig.~\ref{fig:coupled-Fano} we compare  the spectra of absorbed power obtained from Eq.~(\ref{eq:x})-(\ref{eq:dark}) with the measured extinction spectra.
The spectra of absorbed power $P_x(\omega_s)$ of both isolated and coupled dipoles $x=[u,v,\zeta]$ are calculated from the relation $x(t)=x(\omega_s)e^{i\omega_st}$, $P_x(t)=F_x(t)\dot{x}(t)$.
When near-field couplings are absent, i.e., $f=g=0$, the red, blue, and green curves in Fig.~\ref{fig:coupled-Fano}(a) correspond to the absorbed powers in the isolated Lorentzian oscillators with resonances at $\omega_L$ and $\omega_H$ of DSRR in MM${\bf 1}$ and $\omega_M$ of FRR
in MM${\bf 2}$, respectively,
We set the parameters as shown in Table~\ref{table-isolate} for resonance frequency in {\rm sec}$^{-1}$, damping $\gamma$ in {\rm sec}$^{-1}$, and quality factor Q of isolated Lorentzian oscillators $u(t)$, $v(t)$, and  $\zeta(t)$ of Fig.~\ref{fig:coupled-Fano}(a). Note that we adopt the Lorentzian resonance ${\omega_{M}}^{(1)} = 60 {\rm cm}^{-1}$ and ${\gamma_{M}}^{(1)} = 14 {\rm cm}^{-1}$ with $Q^{(1)} = 4.4 $ in Fig.~\ref{fig:S3} as the subradiant mode $\zeta(t)$.
The quality factor Q of $\omega_M$ oscillator of FRR, the isolated $\zeta(t)$ dipole, is the highest, asserting that $\omega_M$ oscillator is subradiant.
%

%

\begin{table}[t]
\centering\caption{Resonance frequency $\omega$ in {\rm sec}$^{-1}$, damping $\gamma$ in {\rm sec}$^{-1}$, and quality factor $Q$ of isolated Lorentzian oscillators $u(t)$, $v(t)$, and  $\zeta(t)$ of Fig.~\ref{fig:coupled-Fano}(a)
in the absence of near-field couplings}
\begin{tabular}{>{\centering\arraybackslash}
p{0.3in}p{0.3in}p{0.25in}|p{0.3in}p{0.3in}p{0.25in}|p{0.3in}p{0.3in}p{0.25in}}
\hhline{===|===|===}
{\it u(t)},& red & &
{\it v(t)},& blue& &
${\it \zeta(t)}$, & green&  \\
\hhline{===|===|===}
$\omega_u$&$\gamma_u$& $Q_u$ &
$\omega_v $&$\gamma_v$ & $Q_v$ &
$\omega_\zeta$&$\gamma_\zeta$ & $Q_\zeta$ \\
\hline
47&15&3.1  &
70&25&2.8  &
60 & 14 &4.3
\\
\hline
\end{tabular}
{\label{table-isolate}}
\end{table}

\begin{table}[t]
\centering\caption{Coupling strengths and driving amplitude in Eq.~(\ref{eq:x})-(\ref{eq:dark}) in the presence of near-field couplings with $\Omega_r=31 {\rm sec}^{-1}$}
\begin{tabular}{>{\centering\arraybackslash}m{1.2in}
|>{\centering\arraybackslash}m{1.2in}|>{\centering\arraybackslash}m{1.2in}}
\hhline{=|=|=}
{\it f}& {\it g} &$\eta$\\\hline
1.0&1.0&0.1
\\
\hline
\end{tabular}
{\label{table-coupled}}
\end{table}

In the presence of near-field couplings, i.e., $f\ne 0$ and $g\ne 0$, two independent superradiant oscillators $u(t)$ and $v(t)$ residing in DSRR  couple with a common subradiant oscillator $\zeta(t)$ of FRR via near-field couplings. With the parameters listed in Table~\ref{table-coupled}, we obtain the absorbed powers of each coupled oscillator as shown in Fig.~\ref{fig:coupled-Fano}(b).
Gray, pink, and dark brown curves correspond to the absorbed powers of $\omega_L$ and $\omega_H$ oscillators of DSRR and $\omega_M$ oscillator of FRR in MM${\bf 3}$, respectively, in the presence of near-field couplings. Black curve is the sum of gray, pink, and dark brown curves.
Near-field couplings of $\omega_L$ oscillator (red) and $\omega_H$ oscillator (blue) with the subradiant $\omega_M$ oscillator (green) give rise to the characteristic asymmetric Fano resonances of the coupled dipoles in Fig.~\ref{fig:coupled-Fano}(b).
Gray curve of Fig.~\ref{fig:coupled-Fano}(b) shows that there occurs a Fano dip in the high energy side of $\omega_L$ resonance of isolated $u(t)$ (red) of Fig.~\ref{fig:coupled-Fano}(a).
Pink curve of Fig.~\ref{fig:coupled-Fano}(b) shows that there occurs a Fano dip in the low energy side  of $\omega_H$ resonance of isolated $v(t)$ (blue) of Fig.~\ref{fig:coupled-Fano}(a). While the dipolar excitations in $\omega_L$ oscillator and $\omega_H$ oscillator are independent, the resulting double Fano resonances in the black curve show that the decrease and increase in the absorbed powers of DSRR are correlated by the common $\omega_M$ oscillator.

For comparison, in Fig.~\ref{fig:coupled-Fano}(c) and (d), we plotted the measured spectra of MM${\bf 1}$ and  MM${\bf 2}$ in purple and green curves, respectively, along with MM${\bf 3}$ spectrum in black curve.  We find that the relation of MM${\bf 1}$ $\&$ MM${\bf 2}$ spectra and MM${\bf 3}$ spectrum is well described by the coupled equations of motion Eq.~(\ref{eq:x})-(\ref{eq:dark}).
In other words, in the composite MM${\bf 3}$ possessing tripod plasmonic resonances, the correlation between Fano dip and extinction enhancement stems from an interplay of $\omega_L$ and $\omega_H$ oscillators in double Fano resonances, mediated by coupling $\Omega_r^2$  coherently coupled to the common narrow oscillator.

\subsection{Phases in double Fano resonances}
\begin{figure}[t]
\begin{center}
        \includegraphics[width=14cm]{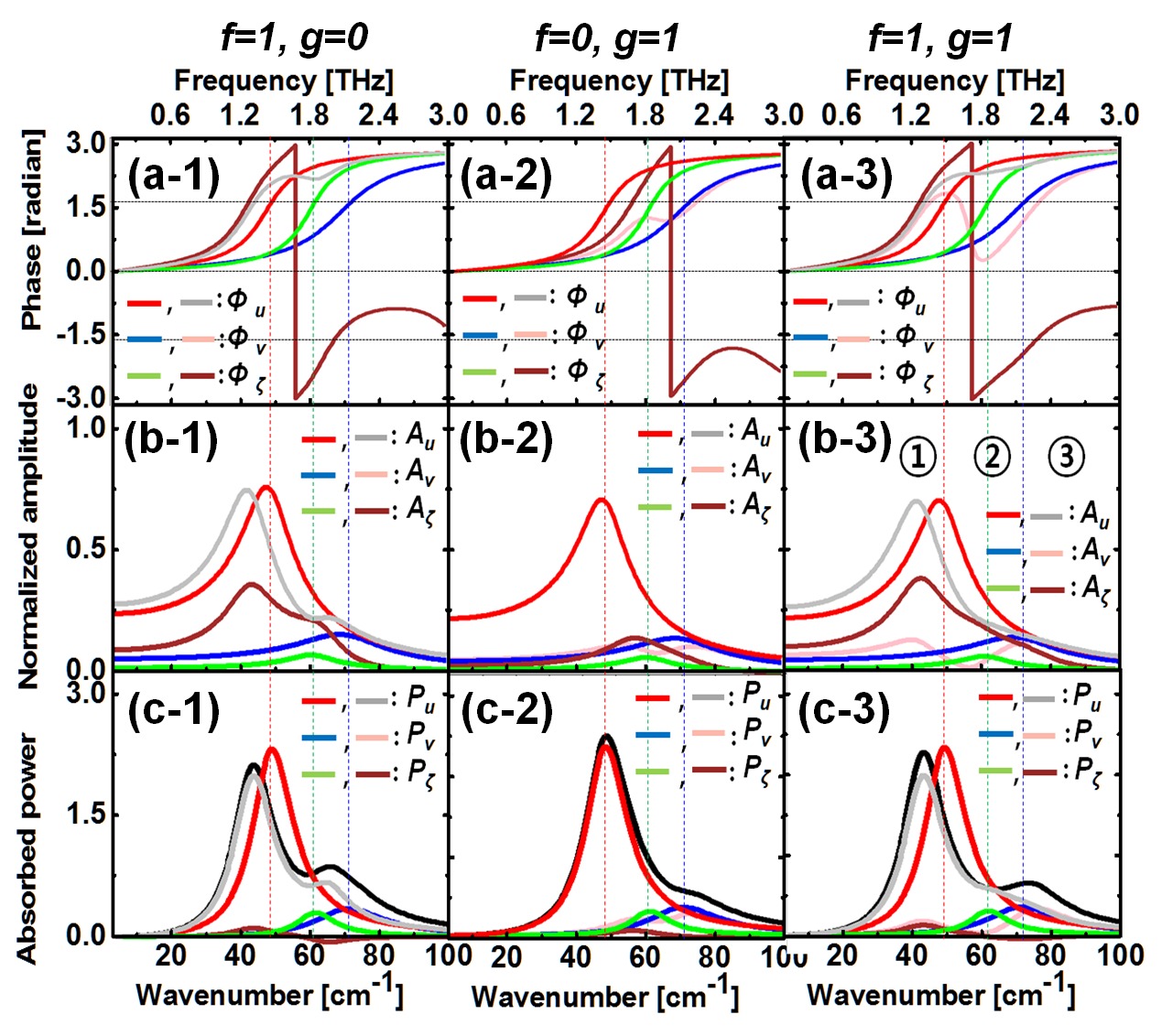}
   \caption
   {\label{fig:phase}
Phase $\Phi_{x}$, amplitude $A_{x}$, and absorbed powers $P_{x}$ are plotted for $f=1, g=0$ (left panel), $f=0, g=1$ (center panel), and $f=1,g=1$ (right panel). In all the figures, as a reference, phase $\Phi_{x}$, amplitude $A_{x}$, and absorbed power $P_{x}$ of isolated dipoles are plotted in the absence of near field couplings, $f=0,~g=0$, in red, blue, and green curves. See text for a detailed description.
}
   \end{center}
\end{figure}

In Fano resonance, a destructive interference between superradiant and subradiant modes can be understood by examining how phases of superradiant and subradiant modes evolve upon increasing the driving frequency\cite{joe2006classical}. In the case that the resonance frequency of subradiant mode is higher than that of superradiant mode, the superradiant mode experiences a phase of $\frac{\pi}{2}$ when going through its own superradiant resonance and the phase keeps on increasing, and when passing through the subradiant resonance the phase experiences a rapid decrease to the minimum phase of a value less than $\frac{\pi}{2}$ and recovers back to its original value, eventually accumulating to a total value of $\pi$. The phase evolution of subradiant mode, however, is very different.
The subradiant mode does not experience a decrease in the phase but keeps on accumulating the phase to a total value of $2\pi$ by going through the superradiant resonance and its own subradiant resonance.
Hence near the subradiant resonance there occurs a spectral region where superradiant and subradiant modes oscillate out-of-phase, which gives rise to a destructive interference for Fano resonance.

When double Fano resonances take place in tripod plasmonic resonances, on the other hand, two superradiant modes are coupled to the common subradiant mode, and the phase evolutions of two superradiant modes are not independent of each other but are correlated in a way determined by the relative spectral positions of the superradiant resonances with respect to the subradiant resonance.

In Fig.~\ref{fig:phase} we closely examine phase $\Phi_{x}$ and amplitude $A_{x}$ of dipole oscillators comparing with the absorbed powers $P_{x}$, where $\Phi_{x}$=$\arctan[\textrm{Im}(x(\omega_s))/\textrm{Re}(x(\omega_s))]$, $A_{x}$=$\mid x(\omega_s)\mid$, and $P_{x}$=$f_{x}\textrm{Re}(-i\omega_{s}x(\omega_s))$, where $f_{x}$ is an external force driving $x(t)$ dipole oscillator, $x=[u,v,\zeta]$.
Figure~\ref{fig:phase}(a-1), (a-2), and (a-3) show phase $\Phi_{x}$, and Fig.~\ref{fig:phase}(b-1), (b-2), and (b-3) show amplitude $A_{x}$, and Fig.~\ref{fig:phase}(c-1), (c-2), and (c-3) show absorbed power $P_{x}$ of $x(t)$ dipole oscillator, $x=[u,v,\zeta]$.
In all the figures, as a reference, we show phase $\Phi_{x}$, amplitude $A_{x}$, and absorbed power $P_{x}$ of isolated dipoles in the absence of near field couplings, $f=0,~g=0$, in red, blue, and green curves. Red and blue curves correspond to the superradiant isolated dipoles $u(t)$ and $v(t)$, respectively, while green curve corresponds to the subradiant isolated dipole $\zeta(t)$ with the parameters of Lorentzian oscillators shown in Table~\ref{table-isolate}.

We consider 3 different cases of near-field couplings, i.e., $f=1,~g=0$ (left panel), $f=0,~g=1$ (center panel), and $f=1,~g=1$ (right panel).
In the presence of coupling between superradiant modes and subradiant mode red, blue, and green curves in the plots correspond to the isolated dipoles $x(t)$
From a comparison of red and gray curves in Fig.~\ref{fig:phase} (a-1), we find that the phase evolution of $u(t)$ (gray) exhibits a dip near the resonance of  subradiant $\zeta(t)$ (dark brown), a feature characteristic of a single Fano resonance. Comparison of green and dark brown curves shows that the phase of subradiant $\zeta(t)$ (dark brown) increases to $\pi$ when passing the resonance of superradiant $u(t)$ (red) which takes place before its own resonance of $\zeta(t)$ (green).
In Fig.~\ref{fig:phase} (a-2) we observe a similar behavior in the phase evolution of $v(t)$ (pink), but the phase of subradiant $\zeta(t)$ (dark brown) increases to $\pi$ by passing its own resonance of $\zeta(t)$ (green) before it comes across the resonance of superradiant $v(t)$ (blue).
We note that the spectral positions of $\pi$ phase of the subradiant dipole $\zeta(t)$ are located below and above its resonance when coupled to $u(t)$ and $v(t)$, respectively.
It is evident that the phase dips in superradiant $u(t)$ and superradiant $v(t)$ of Fig.~\ref{fig:phase} (a-1) and (a-2) are responsible for single Fano resonance as seen in the extinction spectra of Fig.~\ref{fig:phase} (c-1) (gray) and (c-2) (pink). Black curves are the extinction spectra of MM${\bf 3}$ when a single Fano resonance takes place. That is, black curve is the sum of gray, blue, and dark brown curves in Fig.~\ref{fig:phase} (c-1), and is the sum of red, pink, and dark brown curves in Fig.~\ref{fig:phase} (c-2).

A numerical integration of absorbed powers shows that the sum of areas under red, blue, and green curves is identical to the sum of areas under gray, pink, and dark brown curves. Negative values of the absorbed powers by the coupled dipoles $v(t)$ (pink) and $\zeta(t)$ (dark brown) which has the additional phase shifts stand for a transfer of powers to the other dipole, which is observed in Fano resonance in plasmonic structures\cite{PhysRevLett.108.077404}. However, we note that the amplitudes of the coupled dipoles $v(t)$ and $\zeta(t)$ are always positive-definite.


In double Fano resonances, however, the phases of $u(t)$ and $v(t)$ are not independent of each other since a common subradiant dipole $\zeta(t)$ is coupled to both $u(t)$ and $v(t)$.
In particular, the phase of $v(t)$ (pink) is strongly affected by the presence of $u(t)$, while the phase of $u(t)$ (gray) is not much different from that in single Fano resonance, as can be seen in Fig.~\ref{fig:phase} (a-1), (a-2), and (a-3).
Most importantly, a phase increase to $\pi$ of subradiant $\zeta(t)$ (dark brown) takes place when passing the resonance of superradiant $u(t)$ (red), even though $\zeta(t)$ is also coupled to superradiant $v(t)$ at the same time.
As a result, the phase of $v(t)$ (pink) is pulled down sharply near the resonance of $\zeta(t)$ (green), which leads to a delay in recovering back to its original value, when compared with what happens in a single Fano resonance. See blue and pink curves at the phase $\pi/2$ in Fig.~\ref{fig:phase}(a-2) and (a-3).
%
%
We also examine the amplitude evolution of the superradiant dipole $v(t)$ (pink) in Fig.~\ref{fig:phase} (b-2) and (b-3). We find that the spectral position of amplitude maximum at the low frequency is red-shifted to a frequency below the resonance of superradiant dipole $u(t)$ (red).

These two changes in phase and amplitude of $v(t)$, namely, a delay in phase recovery and a red-shift of amplitude maximum, are mainly responsible for the difference in extinction spectra (black curves) between single Fano resonances of Fig.~\ref{fig:phase} (c-1) \& (c-2) and double Fano resonances of Fig.~\ref{fig:phase} (c-3).
%
%
In Fig.~\ref{fig:phase} (c-3) we can identify how double Fano resonances take place by examining the extinction spectra of isolated $u(t)$ (red), isolated $v(t)$ (blue), and MM${\bf 3}$ (black) where $u(t)$ and $v(t)$ are coupled to $\zeta(t)$ (green).
In region \circona, both enhancement and red-shift of extinction in $u(t)$ as seen in the change from red curve to black curve originate from the red-shift of amplitude maximum of dipole $v(t)$. In region \circonc, on the other hand, enhancement and blue-shift of extinction in $v(t)$ as seen in the change from blue curve to black curve is due to the delay in phase recovery of dipole $v(t)$.
These two changes give rise to a Fano dip wider than that in a single Fano resonance in region \circonb.
These features are examples of coherent effects in double Fano resonances resulting from near-field couplings of two superradiant oscillators $u(t)$ and $v(t)$ to a common subradiant oscillator $\zeta(t)$ in a tripod metamaterial system. In the extinction spectra of Fig.~\ref{fig:phase} (c-3) the coherent effects show up as an energy transfer from low-energy $\omega_L$ oscillator to high-energy $\omega_H$ oscillator in DSRR, which was mentioned in the discussion of Fig.~\ref{fig:coupled-Fano}.

\section{Electric field distributions in double Fano resonances}
\begin{figure}[t]
  \begin{center}
   \includegraphics[width=11cm]{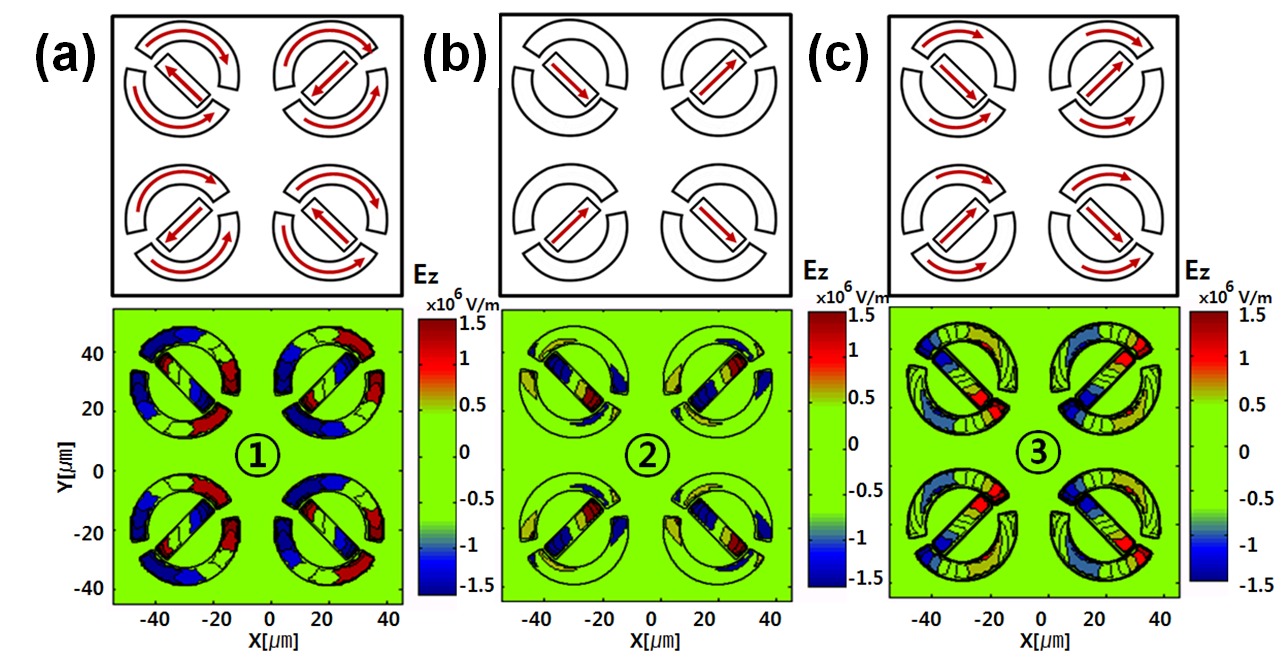}
   \caption
   {\label{fig:E_z-plot}
 $z$-components of near-field electric field, $E_z$, of MM${\bf 3}$ are plotted for a normally incident wave with horizontal  polarization ($x$-axis). (a), (b), and (c) correspond to  regions \circona, \circonb, and \circonc of Fig.~\ref{fig:phase}, respectively. In particular, (b) corresponds to the frequency 1.8THz which is away from the Fano dip.}
    \end{center}
  \end{figure}
   \begin{figure}[t]
  \begin{center}
   \includegraphics[width=11cm]{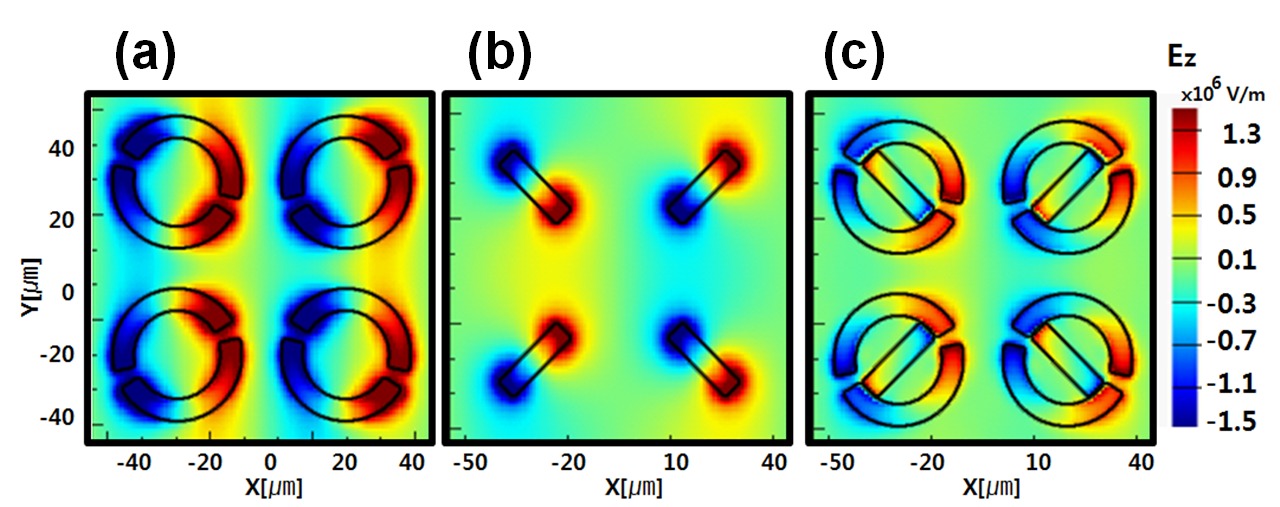}
   \caption
   {\label{fig:three-distributions}
 $z$-components of near-field electric field, $E_z$, of (a) the isolated DSRR, (b) the isolated FRR, and (c) the composite MM${\bf 3}$ are plotted at the frequency 1.6THz, where the Fano dip takes place.
}
    \end{center}
  \end{figure}

One advantage of the coupled equations of motion Eq.~(\ref{eq:x})-(\ref{eq:dark}) in describing
double Fano resonances is that two orthogonal dipolar oscillations $u(t)$ and $v(t)$ residing in DSRR of MM${\bf 3}$ can be identified.
On the other hand, an FDTD simulation provides the information on the electric field distribution of MM${\bf 3}$, which allows us to identify  the electric multipolar nature of coherent coupling between DSRR and FRR in MM${\bf 3}$.

All the FDTD simulation is conducted with a commercial software FDTD solution Lumerical \cite{lumerical}.
We plot distributions of $z$-component of electric fields, which are shown in Fig.~\ref{fig:E_z-plot} for a normally incident wave with horizontal  polarization ($x$-axis).
In a planar metamaterial, it is known that $z$-components of the electric field, normal to the planar metamaterial surface ($x-y$ plane), provide the information on planar charge distributions of the plamonic excitations\cite{gu2012active,shen2012photoexcited,He2013371,Bitzer:11}.
First of all, we note that excitations in DSRR and  FRR of the composite MM${\bf 3}$ are dipolar, similar to dipolar double Fano resonances reported in Ref.~\cite{artar2011directional}, where a single dipolar resonance is coupled two dipoles placed side-by-side, and double Fano resonances are observed in opposite ends of the spectrum with respect to a single dipolar resonance.
In region \circona, dipolar excitation of $\omega_L$ oscillator is dominant in DSRR, and the dipoles in DSRR and FRR are anti-parallel forming a bonding hybridization. In region \circonc, on the other hand, dipolar excitation of $\omega_H$ oscillator is dominant in DSRR, and the dipoles in DSRR and FRR are parallel forming an anti-bonding hybridization.

In region \circonb, a Fano dip takes place at the frequency 1.6THz. In order to show explicitly the excitation of subradiant mode, distributions of z-component of electric fields in the isolated DSRR, the isolated FRR, and the composite MM${\bf 3}$ are shown in Fig.~\ref{fig:three-distributions}.
Comparison of Figs.~\ref{fig:three-distributions}(a) and (c) shows that the induced dipole of DSRR is significantly suppressed by a coherent coupling with FRR, giving rise to the Fano dip in the spectrum of DSRR in MM${\bf 3}$. Most importantly, the induced dipole of FRR in Fig.~\ref{fig:three-distributions}(c) is opposite to that of FRR in  Fig.~\ref{fig:three-distributions}(b).
While the FDTD simulation provides the electric field distribution, thereby, clarifies the multipolar nature of excitations of double Fano resonances, it cannot distinguish two orthogonal superradiant dipolar excitations in DSRR of MM${\bf 1}$ or DSRR of MM${\bf 3}$.

 \begin{figure}[b]
  \begin{center}
     \includegraphics[width=10cm]{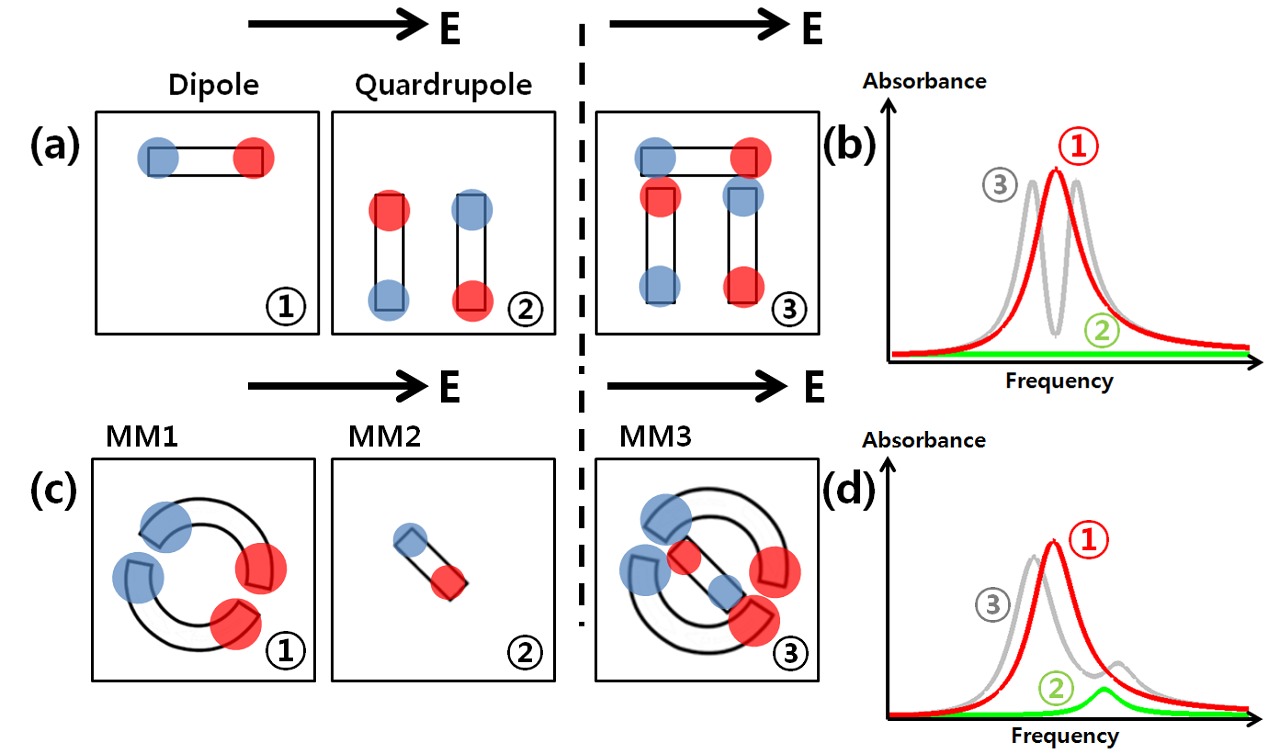}
   \caption
   {\label{fig:schematics-quadrupole-FRR}
 Schematics of asymmetric Fano spectral shape are drawn for (a) dipole-quadrupole and (b) DSRR-FRR structures at the Fano dip.
}
    \end{center}
  \end{figure}

By comparing Fano resonances dipole-quadrupole and DSRR-FRR structures, we clarify the Fano mode of dipole excitations in FRR. Figure~\ref{fig:schematics-quadrupole-FRR} shows schematics of Fano mode excitations.
The left panel represents the uncoupled (a) dipole-quadrupole structure and (c) DSRR-FRR structure. While a quadrupole is excited in two parallel rods in dipole-quadrupole structure, a weak dipole is excited in FRR in DSRR-FRR structure.
A Fano dip in dipole-quadrupole structure comes from a suppression of dipole excitation owing to a coherent coupling with the quadrupole excitation in two parallel rods as shown in Fig.~\ref{fig:schematics-quadrupole-FRR}(b). In DSRR-FRR structure, on the other hand, a dipole excitation in subradiant FRR induced by a near-field coupling with DSRR suppresses a superradiant mode dipolar excitation in DSRR of MM${\bf 1}$ as shown in Fig.~\ref{fig:schematics-quadrupole-FRR}(d).
%

\begin{figure}[t]
  \begin{center}
   \includegraphics[width=9cm]{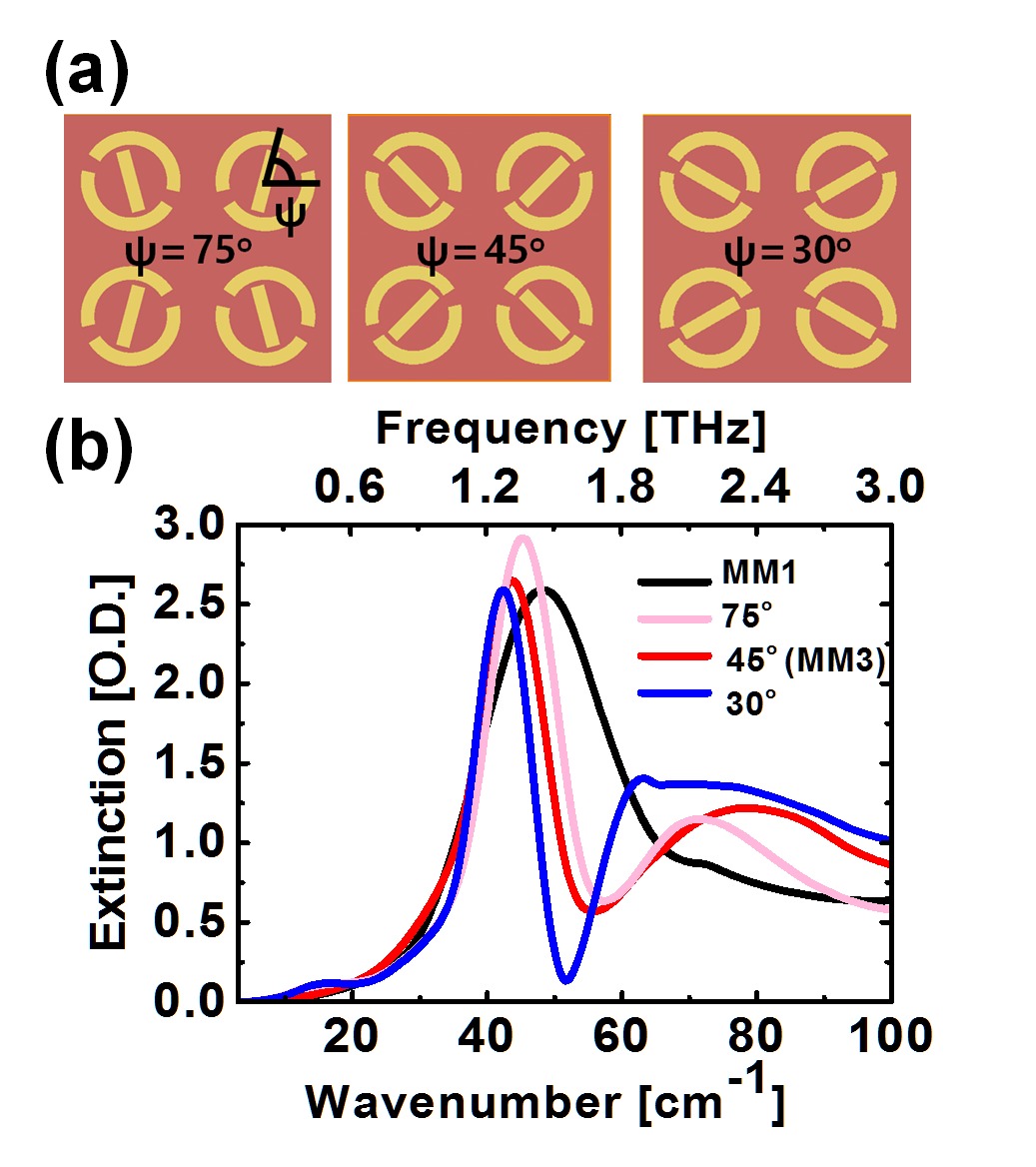}
   \caption {For a series of relative orientations of DSRR and FRR, $\psi = 75^{\circ}$, $\psi = 45^{\circ}$, and $\psi = 30^{\circ}$, extinction spectra of the composite MM${\bf 3}$ are shown for $x$-polarization. (a) show schematics of the composite MM${\bf 3}$ and (b) shows the FDTD simulated extinction spectra for $\psi = 75^{\circ}$ (pink curve), $\psi = 45^{\circ}$ (red curve), and $\psi = 30^{\circ}$ (blue curve) along with that of MM${\bf 1}$.}
   {\label{fig:S4}
}
   \end{center}
\end{figure}

We briefly examine how double Fano resonances behave as the relative orientation of DSRR and FRR changes by FDTD simulation. As shown in Fig.~\ref{fig:S4}, MM${\bf 3}$ with $\psi = 30^{\circ}$ exhibits a deeper Fano dip at the spectral location of a lower energy than MM${\bf 3}$ with  $\psi = 45^{\circ}$, while MM${\bf 3}$ with $\psi = 75^{\circ}$ possesses double Fano resonances similar to MM${\bf 3}$ with $\psi = 45^{\circ}$. We suppose that the highest electric field confinement takes place near edges of DSRR and FRR for MM${\bf 3}$ with $\psi = 30^{\circ}$.

\section{Summary}
By embedding four-rod resonators inside double-split ring resonators superlattice,
a composite metamaterial system is introduced, where a common subradiant oscillator in four-rod resonator is coupled with two superradiant oscillators in double-split ring resonator to give rise to double Fano resonances.
As a classical analogue of four-level tripod atomic system, the extinction spectrum of the composite metamaterial exhibits a double Fano-based coherent effect.
Double Fano resonances are correlated owing to the conjugation of two coupling strengths $f$ and $g$, which shows up as a transfer of absorbed power from one superradiant dipole to the other superradiant dipole in double-split ring resonators.
%
%
%
Double Fano resonances in a tripod metamaterial system can be utilized in designing plasmonic resonances of metamaterial to engineer coherent effects in metamaterials.

\section*{Acknowledgments}
This work is supported by Quantum Metamaterial Research
Center program and Global Frontier Program of Center for Advanced Meta-Materials (Ministry of Science, ICT and Future Planning, National Research Foundation, Republic of Korea).

\newpage

\bibliographystyle{unsrt}

\newpage\vspace{2cm}\noindent

\section*{Appendix: Fano resonance in two coupled oscillators driven simultaneously}

We show that the characteristic asymmetry feature of Fano resonance is retained in plasmonic structure when both superradiant and subradiant oscillators are externally driven.

Originally Fano resonance is introduced to explain an asymmetric peak in excitation spectra observed near autoionization discrete energy level overlapping with continuum energy level of helium.\cite{PhysRev.124.1866}
While the discrete state (dark mode) is not directly excited, a rapid $\pi$ phase change takes place in the phase of a slowly varying-phase background continuum (bright mode) near the discrete resonance. Hence in-phase and out-of-phase interferences take place between continuum and discrete resonances, giving rise to an asymmetric line shape in the continuum spectrum. The sequence of constructive and destructive interferences depends on whether the discrete resonance falls on the red or blue side of the background continuum resonance.
The asymmetric Fano line shape is characterized by 3 parameters, the resonance shift  $\Delta$, the resonance width $\Gamma$, and the asymmetry parameter $q$.

Differently from a dark mode in atomic spectroscopy, there exits an intrinsic non-radiative Joule loss $\gamma_i$ in a plamonic dark mode when  electromagnetically driven, and Gallinet {\it et} {\it al.}~\cite{7.Abinitio} obtained a general formula of Fano-like asymmetric line shape in metallic plasmonic structures, where a modulation damping parameter $b$ is introduced, a measure of the decrease in Fano resonance contrast~\cite{Gallinet:11}.  It is also shown that a classical two coupled oscillators model~\cite{alzar2002classical} with a lossy dark oscillator provides the same general formula when the complex amplitude of the bright oscillator is expanded near the dark resonance~\cite{gallinet2013plasmonic}.

Furthermore, even without resorting to a plasmonic dark mode, a far-field Fano-like asymmetric line shape is observed in plasmonic structures. The occurrence of a Fano-like interference in a single nano-rod plasmonic structures is attributed to a spatial overlap of two non-orthogonal adjacent plasmonic modes~\cite{lopez2012fano}.
And in a nonconcentric ring/disk cavity plasmonic structure, Fano line shape still persists when dark multipolar resonances are optically excited by a retardation effect at a non-normal incidence angle~\cite{hao2008symmetry,hao2008shedding}.
In a heptamer plasmonic nanostructure composed of one center disc with surrounding hexamer ring, an examination of local density of optical states by a selective e-beam excitation of center and ring showed that a far-field interference between the dressed superradiant  and subradiant  eigenmodes fully describes Fano resonance~\cite{PhysRevLett.108.077404}.
In fact, asymmetric Fano line shapes observed in a variety of plasmonic structures are fitted to the general Fano formula with an excellent agreement~\cite{gallinet2011influence}.
These studies lead to the notion that in plasmonic structures an interference between superradiant and subradient eigenmodes overlapping both spectrally and spatially gives rise to an asymmetric Fano line shape resonance~\cite{lopez2012fano,lovera2013mechanisms}.

Now we raise the question whether asymmetric Fano line shape is retained even when both superradiant and subradiant oscillators are driven simultaneously in the presence of a coherent coupling. We examine how the broad Lorentzian spectrum of a superradiant oscillator is modified when coupled with a driven subradiant narrow oscillator by adopting a simple mechanical two coupled oscillators model.
Two coupled oscillators are described by a set of equations of motion.
\begin{eqnarray}
\frac{d^2 x_b}{dt^2} + \gamma_b \frac{dx_b}{dt} + \omega_b^2 x_b + g x_d &=& f_1e^{i\omega t} {\label{eq:b}}\\
\frac{d^2 x_d}{dt^2} +  \gamma_d \frac{dx_d}{dt} + \omega_d^2 x_d + g x_b &=& f_2e^{i\omega t} {\label{eq:d}}
\end{eqnarray}
Narrow spectrum of the subradiant oscillator $d$ is located within broad spectrum of the superradiant oscillator $b$.
Putting $x_b = c_be^{i\omega t}$, we obtain
the amplitude $c_b$ of the superradiant oscillator $b$ near resonance $\omega_d$ of narrow subradiant oscillator $d$.
\begin{eqnarray}
c_{b}(\omega) &=&\frac{(\omega_d^2 + i\gamma_d\omega-\omega^2)f_1-g f_2}{L_\omega(\omega_d^2+i\gamma_d\omega-\omega^2)-g^2}\nonumber\\
&=&\frac{f_1}{L_\omega}\cdot\frac{({\omega_d^2 + i\gamma_d\omega-\omega^2 -g \frac{f_2}{f_1}})L_\omega}{L_\omega(\omega_d^2 + i\gamma_d\omega-\omega^2)-g^2}\nonumber\\
&\approx&\frac{f_1}{L_\omega}\cdot\frac{({\omega_d^2 + i\gamma_d\omega-\omega^2 -g \frac{f_2}{f_1}})L_d}{L_d (\omega_d^2 + i\gamma_d\omega-\omega^2)-g^2}\nonumber
\end{eqnarray}
where $L_\omega\equiv\omega_{b}^2+i\gamma_b\omega-\omega^2$ and  $L_d =\omega_{b}^2+i\gamma_b\omega_d-\omega_d^2$.
An external driving $f_2$ of the subradiant oscillator contributes to the superradiant amplitude via a coupling $g$.
Following procedures in Ref.~\cite{gallinet2013plasmonic},
the superradiant oscillator spectrum has the following relation.
\begin{equation}
\vert c_{b}\vert^2 \approx \frac{f_1^2}{\vert L_\omega\vert^2}\cdot \frac{(\kappa+q_1)^2+b}{\kappa^2+1}\nonumber
\end{equation}
where
\begin{equation}
q_1=q+ \frac{g}{\Gamma} \frac{f_2}{f_1}.\nonumber
\end{equation}
Formula of asymmetric Fano resonance line shape reported in Ref.~\cite{gallinet2013plasmonic} is kept the same as a closed form with a modification in the asymmetry parameter from $q$ to $q_1$.

We observe that a dark mode is not necessarily prerequisite for an asymmetric Fano resonance to take place in plasmonic structures. The dark mode is simply an example of non-driven narrow resonance.
The characteristic asymmetry feature of Fano resonance in plasmonic structures is retained even in the presence of an external driving of narrow oscillator as far as a dispersive coherent coupling between two oscillators ensures a $\pi$-phase shift in the broad superradiant oscillator near the narrow subradiant oscillator resonance.

\end{document}